\documentclass[aps,preprint,showpacs,preprintnumbers,amsmath,amssymb]{revtex4}
\usepackage{amsmath,mathrsfs,amsbsy,color,graphicx,bm,amsthm,amsfonts}
\usepackage{units}
\usepackage{bbm}
\usepackage{times}
\usepackage{dcolumn}
\usepackage{mathrsfs}
\usepackage{amsmath,amssymb,epsfig}

\begin{document}

\title{Monogamy of Einstein-Podolsky-Rosen  steering in the background  of an asymptotically flat black hole}
\author{Jieci Wang$^{1,2}$\footnote{Email: jcwang@hunnu.edu.cn}, Jiliang Jing$^{1}$\footnote{Email:  jljing@hunnu.edu.cn}, and Heng Fan$^{2}$\footnote{ Email: hfan@iphy.ac.cn}}
\affiliation{$^1$ Department of Physics, Synergetic  Innovation Center for Quantum Effects, and Key Laboratory of Low
Dimensional Quantum Structuresand Quantum Control \\
 of Ministry of Education,
 Hunan Normal University, Changsha, Hunan 410081, China\\
 $^2$  Institute of Physics,Chinese Academy of Sciences, Beijing 100190,  China}


\begin{abstract}

We study the behavior of  monogamy deficit and monogamy asymmetry for Einstein-Podolsky-Rosen steering of  Gaussian states under the influence of the Hawking effect. We demonstrate  that the monogamy of quantum steering shows an extreme scenario in the curved spacetime: the first part of a tripartite system cannot individually steer two other parties,  but it can steer the collectivity of the remaining two parties. We also find that the  monogamy deficit of  Gaussian steering,  a quantifier of genuine tripartite steering, are generated due to the influence of the Hawking thermal bath.   Our results elucidate the structure of quantum steering in  tripartite quantum systems in curved spacetime.

\end{abstract}

\vspace*{0.5cm}
\maketitle
\section{Introduction}

 Quantum correlations, including but not limited to entanglement, are considered as key resources for quantum information processing tasks. Einstein-Podolsky-Rosen(EPR)  steering  \cite{reid,wiseman}  is a type of quantum correlation which intrinsically relates to the concept of the EPR paradox  \cite{epr,eprpar}, and was originally realized
by Schr\"odinger \cite{schr2,schr}. Quantum  steering is a quantum phenomenon that allows one to manipulate the state of one subsystem by performing measurements on the other spatially separated subsystem.
 Different from classical correlation, a crucial property of  quantum correlations  is that they
cannot be freely shared among different parties. This property is known as monogamy
 \cite{ibmrev} and  is regarded as one of the most fundamental traits of quantum systems.
Quantitative statements on the monogamy property of quantum entanglement  are due to
Coffman, Kundu, and Wootters (CKW) \cite{ckw}. They proved the following
entanglement monogamy inequality for   a three-qubit state $\rho_{ABC}$:
$
{\cal C}^2_{A:(BC)}\left( \rho_{ABC} \right) \geq {\cal C}^2_{A:B}\left( \rho_{ABC} \right)+{\cal C}^2_{A:C}\left( \rho_{ABC} \right),
$
where ${\cal C}_{A:(BC)}\left( \rho_{ABC} \right)$ is the concurrence of the partition $A$ with
the group $\{BC\}$. 
However, for other forms of quantum correlations much less is known for
quantitative monogamy, let alone the quantum steering which has  been attracted extensive interest only most recently \cite{Skrzypczyk,Walborn,  steering2, steering3, steering4, steering5,steering1,Saunders, Handchen,Sun,steering7,Kocsis,Bowles,Adesso2015,Sabin2015, reidmonogamy,he2016,sun2017}.

According to S. Hawking's finding, a black  hole  would emit
thermal radiation due to quantum effects near the event horizon \cite{Hawking-Terashima}.  This discovery  strongly supports  Bekenstein's phenomenological thermodynamics of black holes. And for this reason, quantum  correlations in fact
intrinsically relevant to the foundational
core of thermodynamics and information loss problem \cite{Hawking-Terashima,Bombelli-Callen} of black holes. On the other hand, the influences of gravity on quantum systems \cite{RQI2, RQI3, RQI6, RQI7, RQI9} cannot be ignored  with the advance in theory and technology of quantum information processing. For example,
it has been experimentally demonstrated that the gravitational effects of  the Earth  notably influence the  precision of atomic clocks for a variation of $0.33$m in height \cite{Alclock}.  Most recently, an experimental test of photonic entanglement in accelerated reference frames  has been reported \cite{RQI8},   in which a genuine quantum state of an entangled photon pair is exposed for different accelerations. However, we have noticed  that the behavior of steering  monogamy is still unclear yet, which is of interest for the entropy and information of  black holes.  Therefore, it is worthwhile to study  the properties of steering monogamy and its asymmetry under the influence of Hawking radiation.

In this paper we  define the steering monogamy deficit to measure the degree of
steering monogamy and the steering monogamy asymmetry to quantitate  the  symmetry of  monogamy.   We mainly seek  answers to the following two questions: how much is the steering monogamy, and  does the monogamy of steering presents some new   properties  in the  curved spacetime? Our model includes three parts, denoted by Alice, Bob and anti-Bob, respectively.  Alice is a Kruskal observer who freely falls into the black hole, while Bob is an accelerated observe who hovers near the event horizon of the black hole. We let the parts observed by Alice  and Bob initially share a two-mode squeezed state with squeezing $s$. In the presence of a black hole, Alice and Bob will not agree  on the definition of  vacuum, that is, the  physical vacuum  in Alice's
coordinates would exist particles in Bob's coordinates. In addition, a hypothetical observer anti-Bob in the interior region of the black hole is embroiled in because Bob is accelerated \cite{Schuller-Mann, adesso3,Dai, RQI4,jieci1,RQI1,sunw2017}.

The outline of the paper is as follows. In Sec. II we give a Gaussian channel description of Hawking effect for  an asymptotically flat black hole. In Sec. III we introduce the Gaussian steering monogamy inequalities, and define the steering monogamy deficit  and the steering monogamy asymmetry.  In Sec. IV we study  behaviors of  the quantum steering monogamy, the steering monogamy deficit,  as well as the steering monogamy asymmetry  in the background of the black hole. The last section is devoted to a brief summary.

\section{Gaussian channel description of Hawking effect for an asymptotically flat black hole  \label{model}}
In this section we introduce the quantum field theory and the Hawking radiation of a bosonic field in the background of an asymptotically flat black hole. We are going to show how the thermal radiation induced by the Hawking effect can be described by a Gaussian  channel \cite{adesso2,wang15}. We consider a massless  bosonic field $\Phi$, which satisfies the Klein-Gordon(K-G)  equation \cite{birelli}
 \begin{eqnarray}
\label{K-G Equation}\frac{1}{\sqrt{-g}}\frac{{\partial}}{\partial
x^{\mu}} \left(\sqrt{-g}g^{\mu\nu}\frac{\partial\Phi}{\partial
x^{\nu}}\right)=0,
 \end{eqnarray}
in a general spacetime.

The line element of an  asymptotically flat black hole, such as the Schwarzschild
black hole,
the Garfinkle-Horowitz-Strominger dilaton black hole \cite{Horowitz}, and the
Casadio-Fabbri-Mazzacurati (CFM) brane black hole \cite{Casadio},  is given by
\begin{eqnarray}\label{metric}
ds^2=f(r)dt^{2}-\frac{1}{h(r)}dr^{2}-R^{2}(r)(d\theta^{2}+\sin\theta^{2}d\varphi^{2}),
\end{eqnarray}
where the parameters $f(r)$ and $h(r)$ vanish at the event horizon
$r=r_{+}$ of the black hole. The surface gravity of
the black hole is given by
$\kappa=\sqrt{f'(r_{+})h'(r_{+})}/2$. By defining the tortoise
coordinates $r_{*}$ as $dr_{*}=dr/\sqrt{f(r)h(r)}$,
the metric (\ref{metric}) can be rewritten as
\begin{eqnarray}\label{new metric}
ds^2=f(r)(dt^{2}-dr_{*}^{2})-R^{2}(r)(d\theta^{2}+\sin\theta^{2}d\varphi^{2}).
\end{eqnarray}  Throughout this paper we set
$G=c=\kappa_{B}=1$.

 Considering  spherical symmetry of the black hole, we express the normal mode solution of the scalar field as \cite{D-R}
\begin{eqnarray}
\Phi_{\omega lm}=\frac{1}{r}\chi_{\omega
l}(r)Y_{lm}(\theta,\varphi)e^{-i\omega t}.
\end{eqnarray}
In this equation  the radial part $\chi_{\omega l}$ obeys the following
 equation
\begin{eqnarray}\label{radial equation}
\frac{d^{2}\chi_{\omega
l}}{dr_{*}^{2}}+[\omega^{2}-V(r)]\chi_{\omega l}=0,
\end{eqnarray}
where
\begin{eqnarray}
\nonumber V(r)&=&\frac{\sqrt{f(r)h(r)}}{R(r)}\frac{d}{dr}\left[\sqrt{f(r)h(r)}\frac{d
R(r)}{dr}\right]\\&+&\frac{l(l+1)f(r)}{R^{2}(r)},
\end{eqnarray}
and $Y_{lm}(\theta,\varphi)$ is the spherical harmonic on the
unit two sphere.

 Solving the K-G equation near the event horizon of the black hole, one can
obtain a set of positive-frequency outgoing modes  $\Phi^{\pm}_{{\Omega},\text{in}}$ and $\Phi^{\pm}_{{\Omega},\text{out}}$,  which  relates to two causally disconnected
regions of spacetime denoted by $in$ (inside the event horizon) and $out$ (outside the event horizon)
\begin{eqnarray}\label{inside mode}
\Phi^+_{{\Omega},\text{in}}\sim e^{i\omega (t-r_{*})},
\end{eqnarray}
\begin{eqnarray}\label{outside mode}
\Phi^+_{{\Omega},\text{out}}\sim  e^{-i\omega (t-r_{*})},
\end{eqnarray}
where  $r_{*}$ is the tortoise coordinate. The modes can be used to quantize
 the scalar field and define the  vacuum state $|0\rangle_S$ in the coordinates of the asymptotically flat black hole. And, for this reason the solutions  in Eqs. (\ref{inside mode})  and (\ref{outside mode}) are usually called  Schwarzschild modes (or Boulware-Schwarzschild modes) \cite{RQI4,jieci1,Fabbri,  Bruschi, Bruschi2}. Since the solutions Eqs. (\ref{inside mode}) and
(\ref{outside mode}) cannot be analytically continued
from the $in$ region to the $out$ region, we must make an analytic continuation and express them in the
Kruskal coordinates \cite{D-R}. The  Kruskal modes can be used to define the Hartle-Hawking vacuum, which corresponds to the Minkowski vacuum in a flat spacetime. In this way the  scalar field  can be quantized in the Schwarzschild  and Kruskal modes respectively, and  the Bogoliubov transformations \cite{ birelli}  are obtained to relate the modes.

However, as the inertial observer Alice is freely to create excitations in any accessible
modes \cite{wang15,Bruschi}, we have to choose Alice's modes as superpositions of
different frequencies of Schwarzschild modes. In other words, it is improper to map a single-frequency  Kruskal mode
into a single-frequency Schwarzschild  mode.  We therefore employ the intermediate Unruh modes, which are positive-frequency combinations of modes in Kruskal coordinates \cite{Bruschi}. Since the vacuum of Unruh and Kruskal modes coincide,  a
positive Kruskal mode can be written as a linear combination
of two Unruh modes as $
c_{\Omega,\text{U}}=q_{\text{R}}C_{\Omega,\text{R}}
+q_{\text{L}}C_{\Omega,\text{L}}$, where
\begin{align}\label{Unruhop}
\nonumber C_{\Omega,\text{\text{R}}}=&\left(\cosh r_{\Omega}\, \hat{a}_{{\Omega},\text{out}}-\sinh r_{\Omega}\, \hat{b}^\dagger_{{\Omega},\text{in}}\right),\\*
 C_{\Omega,\text{\text{L}}}=&\left(\cosh r_{\Omega}\, \hat{a}_{{\Omega},\text{in}}-\sinh r_{\Omega}\, \hat{b}^\dagger_{{\Omega},\text{out}}\right),
\end{align}
and $q_\text{\text{R}},q_\text{\text{L}}$  satisfy $|q_\text{\text{R}}|^2+|q_\text{\text{L}}|^2=1$. In Eq. (\ref{Unruhop}) the Hawing temperature parameter $r_{\Omega}$ is defined as $\sinh r_{\Omega}=(e^{\frac{ 2\pi\Omega}{\kappa}}-1)^{-\frac{1}{2}}$. Here $\kappa$ is the surface gravity of the  asymptotically flat black hole which relates the Hawking temperature $T$ by $T=\kappa/2\pi$.
By employing the Bogliubov transformation between the Unruh modes and the Schwarzschild modes, the Unruh vacuum is found to be   \cite{wang15,Bruschi,Fabbri,Bruschi2}
\begin{equation}\label{vacuumba}
|0_\Omega\rangle_\text{U}=\frac{1}{\cosh r_\Omega^{2}}\sum_{n,m=0}^{\infty}\tanh r_\Omega^{n+m}|mnm'n'\rangle_{\Omega},
\end{equation}
where $|mnm'n'\rangle=|m_{\Omega}\rangle^{+}_{out}
|n_{-\Omega}\rangle^{-}_{in}
|m'_{-\Omega}\rangle^{-}_{out}
|n'_{\Omega}\rangle^{+}_{in}$, and the superscripts $\{+,-\}$  on the kets
is used to indicate the particle and antiparticle modes, respectively. In Eq. (\ref{vacuumba}), the Schwarzschild modes  $\{|n\rangle_{out}\}$ and
$\{|n\rangle_{in}\}$  are observed by Bob who hovers outside the event horizon and  anti-Bob who is a hypothetic observer  inside  the
black hole, respectively. 
The mode parameter can be  fixed as $q_R=1$ and $q_L=0$ by assuming  Bob's
detector  only sensitive to the particle modes and anti-Bob's detector only detects antiparticle modes. In this situation,  Eq.~(\ref{vacuumba}) reduces to $|0_\Omega\rangle_\text{H}=\frac{1}{\cosh r_\Omega}\sum_{n=0}^{\infty}\tanh r_\Omega^{n}|nn\rangle_{\Omega}$, which is a two-mode squeezing state and the squeezing $r_\Omega$ is directly  related  to the Hawking temperature by $ r_{\Omega}=arcsinh[(e^{\frac{ \Omega}{T}}-1)^{-\frac{1}{2}}]$, where $T=\frac{\sqrt{f'(r_{+})h'(r_{+})}}{4\pi}$. Therefore,  the effect of Hawking radiation can be described by a
two-mode squeezing operator $\hat{U}_{out,in}(r_\Omega)$ acting on the input state  $|\psi_0\rangle$
  \begin{eqnarray}\label{stator2}
 \nonumber\rho_{out} &=& {\rm tr}_{in} \{\hat{U}_{out,in}(r_\Omega) \big[(|\psi_0\rangle\!\langle\psi_0|)_{out} \\&& \otimes \ (|0\rangle\!\langle0|)_{in}\big] \hat{U}_{out,in}^\dagger(r_\Omega)\}\,,
 \end{eqnarray}
where the squeezing operator has the form $\hat{U}_{out,in}(r_\Omega)=e^{r_\Omega(\hat{b}^\dagger_{{\Omega},\text{out}}\hat{b}^\dagger_{{\Omega},\text{in}}-
\hat{a}_{{\Omega},\text{out}}\hat{a}_{{\Omega},\text{in}})}$ and   we rewrite $r_\Omega$ as $r$ hereafter.  In this paper we work in the phase space, then we employ a symplectic phase-space  representation $ S_{B,\bar B}(r)$ for the two-mode squeezing transformation, which is  \cite{ adesso3, adesso2,wang15}

\begin{eqnarray}\label{cmtwomode}
 S_{B,\bar B}(r)=\left(
                   \begin{array}{cc}
                     \cosh r I_2 & \sinh r Z_2 \\
                     \sinh r Z_2 & \cosh r I_2 \\
                   \end{array}
                 \right),
\end{eqnarray}
where $I_2$ is  a $2\times 2$ identity matrix and  $Z_2=\left(
                       \begin{array}{cc}
                         1 & 0 \\
                         0 & -1 \\
                       \end{array}
                     \right)$.

\section{Gaussian steering monogamy inequalities,   steering monogamy deficit,  and steering monogamy asymmetry  \label{GSteering}}

In this section we briefly introduce some concepts of the monogamy of Gaussian quantum steering and define the  steering monogamy deficit and the  steering monogamy asymmetry.  For our purposes, we consider a tripartite Gaussian state $\rho_{ABC}$ with covariance matrix $\sigma_{ABC}$.
The elements ${\sigma _{ij}} (i,j=1....6)$
 for the covariance matrix $\sigma_{ABC}$ of state $\rho_{ABC}$ are defined as  ${\sigma _{ij}} = {\rm tr} \big[ {{{\{ {{{\hat R}_i},{{\hat R}_j}} \}}_ + }\ {\rho_{ABC}}} \big]$, where
$\hat R = (\hat x_1,\hat p_1, \hat x_2, \hat p_2,\hat x_3, \hat p_3)^{\sf T}$ is the vector collecting position and momentum operators of each mode, satisfying canonical commutation relations $[{{{\hat R}_i},{{\hat R}_j}} ] = i{(\Omega_{ABC})_{ij}}$, with $(\Omega_{ABC})  = \omega^{\oplus 3}$ and $\omega =  {{\ 0\ \ 1}\choose{-1\ 0}}$ \cite{weedbrook}.   For the Gaussian state  $\sigma_{ABC}$, its monogamy holds if and only if the quantum steering  obeys the following  CKW-type  inequalities  \cite{renyi,he2016},
\begin{align}\label{monogtrip1}
 &{\cal G}^{AB\rightarrow C} \left( \sigma_{ABC} \right)\geq  {\cal G}^{A \rightarrow C}\left( \sigma_{ABC} \right) + {\cal G}^{B \rightarrow C}\left( \sigma_{ABC} \right), \\
 \label{monogtrip2}
 & {\cal G}^{C \rightarrow AB} \left( \sigma_{ABC} \right) \geq {\cal G}^{C \rightarrow A}\left( \sigma_{ABC} \right) + {\cal G}^{C \rightarrow B} \left( \sigma_{ABC} \right).
\end{align}
where the ${2 \to 1}$ Gaussian quantum steering ${\cal G}^{xy\to z}(\sigma_{xyz})$ describes how the collectivity $\{xy\}$ can steer the part $z$,  and $x,y,z$  denotes all the different permutations of $A$, $B$ and $C$.
Similarly, the ${1 \to 2}$ Gaussian ${\cal G}^{z \to xy}(\sigma_{xyz})$  steerability can be obtained by swapping the roles of $\{xy\}$ and $z$, which measures the steerability  from $z$ to the collectivity $\{xy\}$.

In Eqs. (1) and (2),   the  $x \to y$ Gaussian steering is defined as \cite{Adesso2015}
\begin{eqnarray}
\nonumber {\cal G}^{x \to y}(\sigma_{xy})
&:=&
\max\bigg\{0,\,-\sum_{j:\bar{\nu}^y_j<1} \ln(\bar{\nu}^y_j)\bigg\}\\ &=& \max\big\{0,\, {\cal S}(x) - {\cal S}(\sigma_{xy})\big\}\,, \label{GS1}
\end{eqnarray}
and  $\{\bar{\nu}^B_{j}\}$ are  symplectic eigenvalues of the Schur complement of the $x$ part in the covariance matrix $\sigma_{xy}$ \cite{Adesso2015} and ${\cal S}(\sigma) = \frac12 \ln( \det \sigma)$ is  the R\'enyi-$2$ entropy  \cite{renyi}.

In particular,  it is well known that the Gaussian steering measure $\cal G$ is asymmetry, hence there are two kinds of CKW-type monogamy inequalities for quantum steering. In addition,  \emph{ the} monogamy of steering is  very different from the  monogamy of entanglement where only one monogamy inequality is required to be satisfied. To make a quantitative research on the  monogamy of Gaussian steering, here we define two types of monogamy deficits for the   Gaussian steering
\begin{align}\label{restrip1}
 &{\cal D}^{xy : z}= {{\cal G}^{xy \rightarrow z}  - {\cal G}^{x \rightarrow z} - {\cal G}^{y \rightarrow z}}, \\
 \label{restrip2}
 & {\cal D}^{x: yz} ={{\cal G}^{x \rightarrow yz} - {\cal G}^{x \rightarrow y} - {\cal G}^{x \rightarrow z}},
\end{align}
where $x,y,z$  denotes all the different permutations of $A$, $B$ and $C$. For the sake of discussion, we name ${\cal D}^{xy : z}$ as ${2 \to 1}$ steering  deficit and ${\cal D}^{z: xy}$ as ${1 \to 2}$ steering deficit, respectively.
It is worth noting that the monogamy deficits is in fact a meaningful quantitative indicator of genuine tripartite quantum  steering for  Gaussian states \cite{he2016,steering1}. This is because a non-zero  monogamy deficits in all the three steering directions  $\{xy\} \to z$,  $\{xz\} \to y$ and $\{yz\} \to x$ of a tripartite state certifies a sufficient requirement to violate the corresponding biseparable structure of the state.

The above two monogamy deficits for Gaussian steering may be different because the asymmetry of quantum steering itself. We wonder whether the monogamy deficits  is invariant for different steering directions.  Therefore, we define the
steering monogamy asymmetry to measure the degree of
monogamy asymmetry, which is
\begin{eqnarray} \label{residual2}
 {\cal D}^{xy:z}_\Delta =| {\cal D}^{xy : z}-{\cal D}^{z :xy}|,
\end{eqnarray}
By defining the monogamy asymmetry, we in fact demonstrate a relation which connects two types of monogamy deficits and monogamy inequalities. In this work the third part is observed by  the hypothetical observer Anti-Bob in the interior region of the black hole, we  therefore denote the third mode as $\bar B$ rather than $C$.

\section{ Dynamics of  monogamy deficit and  monogamy asymmetry for quantum steering in curved spacetime \label{tools}}

\begin{figure*}[tbp]
\centerline{\includegraphics[width= 13cm]{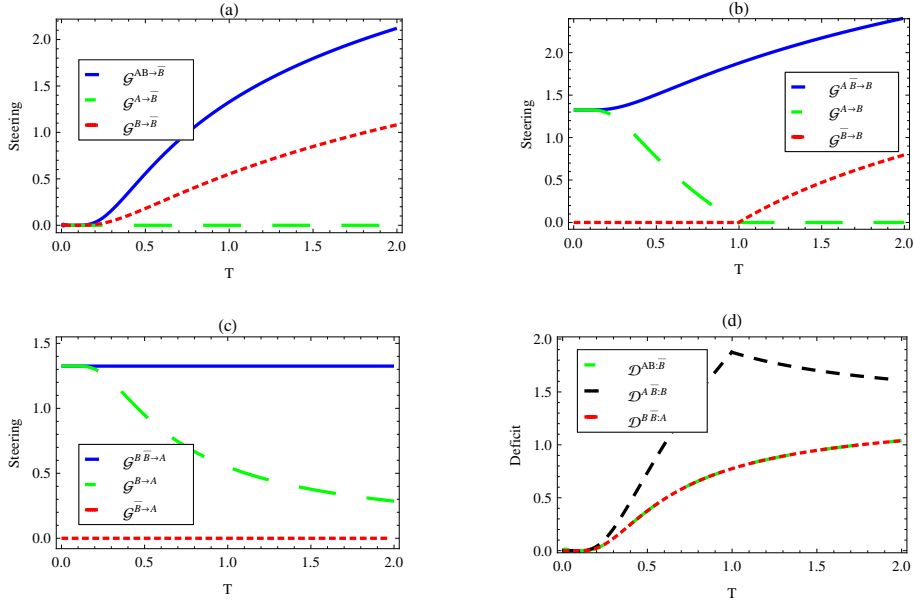}}
\caption{ (a-c) The  ${2 \to 1}$ quantum  steering  (solid lines) and ${1 \to 1}$ quantum  steering (dashed lines) as a function of the Hawking temperature  $T$.   (d) The ${2 \to 1}$ Gaussian steering monogamy deficits ${\cal D}^{AB:\bar B}$, ${\cal D}^{A\bar B:B}$ and ${\cal D}^{B\bar B:A}$ versus   Hawking temperature  $T$. The squeezing parameter $s$ of the initial state is fixed as $s=1$ and  $\Omega$ is fixed as $\Omega=1$.}
\label{Fig1}
\end{figure*}

We consider a massless scalar field $\Phi$  whose state is initialized in a  two-mode Gaussian squeezed state with squeezing $s$. The initial system is prepared in  Unruh modes in the inertial frame with the covariance matrix
\begin{eqnarray}\label{inAR}
\sigma^{\rm (M)}_{AB}(s)=\left(
                           \begin{array}{cc}
                             \mathcal{A}_i(s) & \mathcal{E}_i(s) \\
                             \mathcal{E}^T_i(s) & \mathcal{B}_i(s) \\
                           \end{array}
                         \right),
\end{eqnarray}
where $\mathcal{A}_i(s)=\mathcal{B}_i(s)=\cosh (2s)I_2$,  and $\mathcal{E}_i(s)=\sinh (2s)Z_2$. Because Alice is freely falling into the black hole,
she sees nothing special at the horizon and accesses to
the entire spacetime. However, for the accelerated
observer Bob, an extra set of modes $\bar{B}$,  which is observed  by a  hypothetical observer anti-Bob inside the event horizon, becomes relevant. The transformation of the state from the Unruh modes to the Schwarzschild modes is described by Eq.~(\ref{stator2}). After the transformation, the final state  of the entire three-mode system is given by the  covariance matrix \cite{adesso3}
\begin{eqnarray}\label{in34}
\nonumber\sigma^{\rm }_{AB \bar B}(s,r) &=& \big[I_A \oplus  S_{B,\bar B}(r)\big] \big[\sigma^{\rm (M)}_{AB}(s) \oplus I_{\bar B}\big]\\&& \nonumber\big[I_A \oplus  S_{B,\bar B}(r)\big]\\
 &=& \left(
       \begin{array}{ccc}
          \mathcal{\sigma}_{A} & \mathcal{E}_{AB} & \mathcal{E}_{A\bar B} \\
         \mathcal{E}^{\sf T}_{AB} &  \mathcal{\sigma}_{B} & \mathcal{E}_{B\bar B} \\
         \mathcal{E}^{\sf T}_{A\bar B} & \mathcal{E}^{\sf T}_{B\bar B} &  \mathcal{\sigma}_{\bar B} \\
       \end{array}
     \right)
 \,.
\end{eqnarray}
In Eq. (\ref{in34}) the diagonal elements are $\mathcal{\sigma}_{A}=\cosh(2s)I_2$, $\mathcal{\sigma}_{B}=[\cosh(2s) \cosh^2(r) + \sinh^2(r)]I_2$,  and $\mathcal{\sigma}_{\bar B}=[\cosh^2(r) + \cosh(2s) \sinh^2(r)]I_2$. The non-diagonal elements have the following forms: $\mathcal{E}_{AB}=[\cosh(r) \sinh(2s)]Z_2$, $\mathcal{E}_{B\bar B}=[\cosh^2(s) \sinh(2r)]Z_2$, and $\mathcal{E}_{A\bar B}=[\sinh(2s) \sinh(r)]Z_2$.

\subsection{ The  ${2 \to 1}$ steering monogamy and its deficits   \label{sub1}}
Now let us seek an answer for the first question: is steering  monogamous in the background
of the asymptotically flat black hole?
 We start with the ${2 \to 1}$ monogamy inequality in the ${AB \to \bar B}$ direction. To this end, the  ${2 \to 1}$  quantum steering $ {\cal G}^{AB \to \bar B}$, the ${1 \to 1}$ steering $ {\cal G}^{A \to \bar B}$ and $ {\cal G}^{B \to \bar B}$ are required. After some calculations, we
obtain  
 \begin{eqnarray}
 {\cal G}^{AB \to \bar B} =
\mbox{$\max\big\{0,\,  \ln [{\cosh^2(r) + \cosh(2 s) \sinh^2(r)}]\big\}$},\label{GS21a}
\end{eqnarray}
which  quantifies to what extent part $\bar B$
can be steered by the measurements performed by the collectivity $\{AB\}$. The
 analytic expression of the $A \to \bar B$ and $B \to \bar B$ Gaussian  steering are found to be
 \begin{eqnarray}\label{GS21b}
 \nonumber  {\cal G}^{A \to \bar B} &=&
\mbox{$\max\big\{0,\,  \ln {\frac{\cosh(2s)}{\sinh^2(r) + \cosh(2 s) \cosh^2(r)}}\big\}$},\\
{\cal G}^{B \to \bar B}& =&
\mbox{$\max\big\{0,\,  \ln [{\cosh^2(r) + \frac{\sinh^2(r)}{\cosh(2 s)}]}\big\}$}.
\end{eqnarray}
From  Eqs. (\ref{GS21a}-\ref{GS21b}) we can see that  all the bipartite  Gaussian  steering depends on the Hawking temperature  parameter $r$, which shows that these direction of steerability are affected by thermal noise of  the Hawking radiation.  Substituting Eqs. (\ref{GS21a}-\ref{GS21b}) into
Eq. (9) we find that the  ${1 \to 2}$ steering $ {\cal G}^{AB \to \bar B} $ is always more than the sum of the   ${1 \to 1}$ steering $ {\cal G}^{A \to \bar B}$ and ${\cal G}^{B \to \bar B}$.  That is,  quantum steering in the $\{AB\}\to \bar B$ direction is  monogamous under the influence of the Hawing thermal noise. To better understand the  ${2 \to 1}$ steering  monogamy in the studied curved spacetime, we calculate two other sets of  quantum steerings $\{ {\cal G}^{A\bar B \to  B}, {\cal G}^{A \to B},{\cal G}^{\bar B \to  B} \}$ and $\{ {\cal G}^{B\bar B \to  A}, {\cal G}^{B \to A},{\cal G}^{\bar B \to  A} \}$  and plot them in Fig. (1a-1c), respectively. Furthermore, to seek some quantitative information of the steering monogamy, we also plot the behaviors of steering monogamy  deficits ${\cal D}^{A B:\bar B}$, ${\cal D}^{A\bar B: B}$ and ${\cal D}^{B\bar B: A}$ in Fig. (1d).

\begin{figure*}[tbp]
\centerline{\includegraphics[width= 13cm]{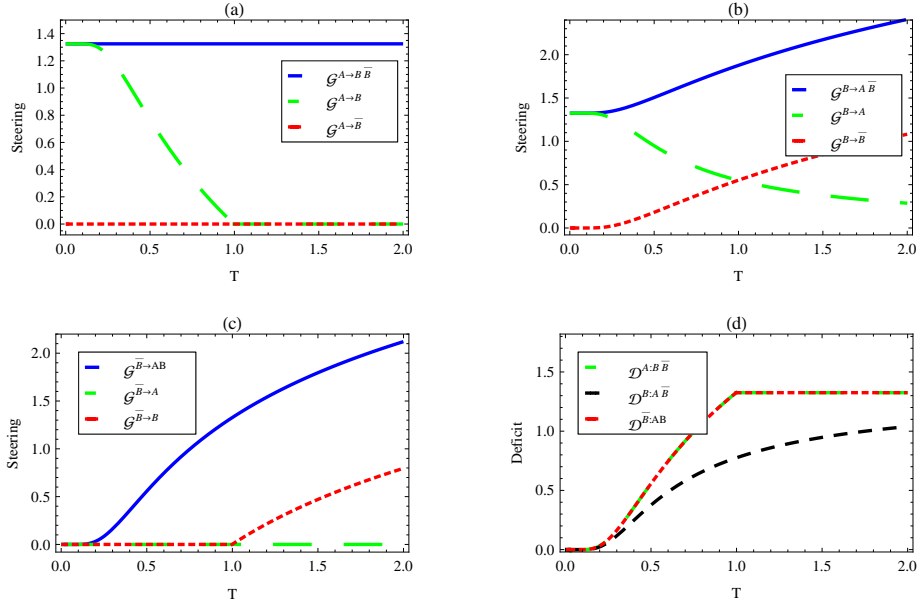}}
\caption{ (a-c) The  ${1 \to 2}$ quantum  steering  (solid lines) and ${1 \to 1}$ quantum  steering (dashed lines) as a function of the  Hawking temperature  $T$.   (d) The ${1 \to 2}$ Gaussian steering monogamy deficits ${\cal D}^{A:B\bar B}$, ${\cal D}^{B:A\bar B}$ and ${\cal D}^{\bar B:AB}$ versus $T$. The squeezing parameter $s$ of the initial state is fixed as $s=1$  and  frequency of the field is fixed as $\Omega=1$.}
\label{Fig2}
\end{figure*}

Fig. (1a) shows how the bipartite quantum  steering $ {\cal G}^{AB \to \bar B} $,   $ {\cal G}^{A \to \bar B}$ and ${\cal G}^{B \to \bar B}$  are influenced by the Hawking radiation of the asymptotically flat black hole. We can see that all of them equal to zero when the Hawking temperature equals to zero, which is because the initial steerability is only prepared between parts $A$ and $B$. As the Hawking temperature increase, quantum steerabilities are created between the assemblies $AB \to \bar B$, $A \to \bar B$ and $B \to \bar B$. However, we can see from Fig. (1b) and Fig. (1c) that the steering $A\to B$ and $B \to A$  always decrease as the  Hawking temperature  $T$ increase. Such a phenomenon indicates the fact that  quantum steerability are distributed among different parts due the influence of Hawking radiation.

 We wonder if the quantum steering can be freely distributed among the system. This question can be answered by check if the ${2 \to 1}$ steering  monogamy holds in this situation. It is shown in Fig. (1d) that  all the three types of ${2 \to 1}$ monogamy  deficits are positive for any Hawking temperature, which means  that the  ${2 \to 1}$ CKW-like  steering  inequalities defined in Eq. (\ref{monogtrip1})  always hold in the curved spacetime. In other words,   the ${2 \to 1}$  quantum steering can not be freely distributed among the system in the curved spacetime.   It is shown  Fig. (1b) that the $A \to B$ steering decreases quickly  and  experiences  ``sudden death", while the $\bar B \to B$ steering appears ``sudden birth" with increasing  $T$. Moreover, the ``sudden death" point of the former is  in accordance with the ``sudden birth" point of the latter. This is also a  powerful evidence for the monogamy of Gaussian quantum steering  in the curved spacetime.

\subsection{ The  ${1 \to 2}$ steering monogamy and its deficits \label{sub2}}

To make a general conclusion for the  steering  monogamy in the  background of a black hole,  next we study the ${1 \to 2}$ steering monogamy inequality given by Eq. (\ref{monogtrip2}) and the ${1 \to 2}$ monogamy deficits defined in Eq. (\ref{restrip2}). The ${A \to\bar B}$ steering has been given in Eq. (18) and the remaining two steerabilities are
 \begin{eqnarray}\label{GS22b}
 \nonumber {\cal G}^{A \to B\bar B} &=&
\mbox{$\max\big\{0,\,  \ln  \cosh(2 s) \big\}$},\\
{\cal G}^{A \to B}& =&
\mbox{$\max\big\{0,\,  \ln {\frac{\cosh(2s)}{\cosh^2(r) + \cosh(2 s) \sinh^2(r)}}\big\}$},
\end{eqnarray}
respectively. Two other sets of  Gaussian  steerings $\{ {\cal G}^{B \to A \bar B}, {\cal G}^{B \to A},{\cal G}^{ B \to  \bar B} \}$
and $\{ {\cal G}^{\bar B \to  AB}, \\{\cal G}^{\bar B \to A},{\cal G}^{\bar B \to  B} \}$  can be computed in a similar way. To compare the behaviors of the ${1 \to 2}$ steering monogamy with its ${2 \to 1}$ counterpart, we plot these three sets of  Gaussian  steerings  as a function of the  Hawking temperature  $T$ for fixed $s=1$  in Figs. (2a-2c). We also calculate  the ${1 \to 2}$  deficits ${\cal D}^{A:B\bar B}$, ${\cal D}^{ B: A\bar B}$ and ${\cal D}^{\bar B: AB}$ and plot the behaviors of them in Fig. (2d).

From Figs.  (2a-2c) we can see that  except
the $A \to B$ and  $B \to A$ steering, all other directions of steering increase with  increasing Hawking temperature  $T$. This reveals the fact that the initially  steering-type quantum correlations prepared between part $A$ and $B$ has been distributed among other directions. It is shown that the $x\to yz$ steering is always more than the sum of $x\to y$ steering and $x-z$ steering for all the three sets of directions. Similar with the entanglement monogamy, the steering monogamy inequalities enjoy a very appealing interpretation: the degree of steerability exhibited by the  collectivity can be larger that the sum of the degrees of steerability exhibited by the individual pairs. We have found in the last subsection that the ${2 \to 1}$ quantum steering shared amongdifferent parties exhibits monogamy property. Here we find again the ${1 \to 2}$ steering  monogamy  still holds under the influence of the Hawing thermal noise.

Moreover, it is interesting to note from Fig. (2a) that the monogamy of Gaussian quantum steering  exists a more extreme scenario: party $A$ cannot individually steer the parties $B$  and $\bar B$, i.e., $ {\cal G}^{A \rightarrow  B} =  {\cal G}^{A \rightarrow \bar B} = 0 $, but the collectivity $\{B\bar B\}$  can be steered by the measurements performed on Alice's side because ${\cal G}^{A\rightarrow B \bar B}>0$. Similar situations exists in Figs. (2c), where the party $\bar B$ inside the event horizon of the black hole cannot steer the parties $A$ and $B$  individually for some small  Hawking temperature  $T$, but it can steer the collectivity  $ \{AB\}$.

Fig. (2d) shows that the ${1 \to 2}$ monogamy  deficits ${\cal D}^{A: B\bar B}$, ${\cal D}^{ B: A\bar B}$ and ${\cal D}^{\bar B: AB}$, a quantifier of genuine tripartite steering of the state,  increase with increasing Hawking temperature. This indicates  genuine tripartite steering have been generated due to the thermal bath introduced by Hawking effect. We also note that the ${\cal D}^{ B: A\bar B}$ monogamy deficit is monotonously increased while ${\cal D}^{A: B\bar B}$ and ${\cal D}^{\bar B: AB}$ deficits exist a transition point.  In addition, the monogamy  deficits  have the same  transition point  with the $1-1$  quantum steerabilities  $A \to B$ and  $\bar B \to B$.  We find that  anti-Bob can steer the \emph{ collectivity} $\{AB\}$ for any  $T$, which is different from the  $1-1$   steering   $\bar B \to B$ where steerable   is bigger than a critical point.  Here antiBob can ensemble steer Alice and  Bob even though they are separated by the event horizon, which verifies  the essential connection between  quantum steering and quantum nonlocality.

\subsection{ Monogamy asymmetry and symmetric behaviors of the ${1 \to 2}$ and  ${2 \to 1}$ steering   \label{sub3}}

\begin{figure}
\includegraphics[scale=1.0]{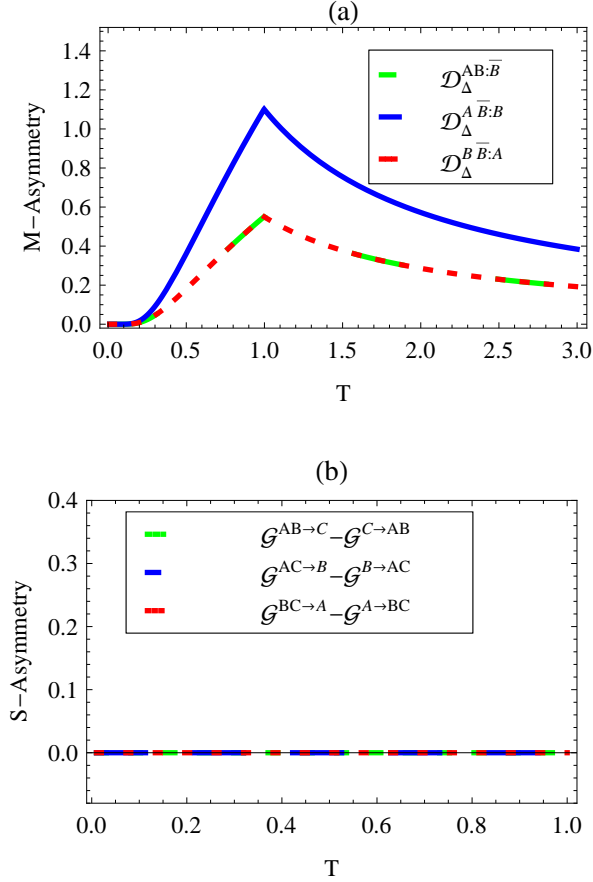}
\caption{(Color online) (a) The monogamy asymmetry of  steering ${\cal D}^{A: B\bar B}_\Delta$, ${\cal D}^{ B: A\bar B}_\Delta$ and ${\cal D}^{\bar B: AB}_\Delta$ for the tripartite  Gaussian state as a function of the Hawking temperature $T$. (b)  The symmetric behaviors of the ${1 \to 2}$  steering and   ${2 \to 1}$ steering versus $T$. The squeezing parameter $s$ of the initial state is fixed as $s=1$  and frequency of the field is fixed as $\Omega=1$.}\label{Fig3}
\end{figure}

Unlike entanglement,  quantum steering is  a type of quantum correlation depending on the direction of measurement \cite{Adesso2015} , which makes the symmetric property of  steering   a crucial issue.  For Gaussian states,  it was recently found that the ${1 \to 1}$  steering is  asymmetric  both in the flat spacetime  \cite{Adesso2015,Sabin2015} and the curved spacetime \cite{wang15}.  We wonder if the asymmetry of steering will  pass to  the monogamy of  steering in the curved spacetime.  To obtain  understanding of this issue, we  calculate  the monogamy asymmetry ${\cal D}^{A: B\bar B}_\Delta$, ${\cal D}^{ B: A\bar B}_\Delta$ and ${\cal D}^{\bar B: AB}_\Delta$ of the Gaussian quantum steerability   and check if the are asymmetric  in the Schwarzschild spacetime.

In Fig. (\ref{Fig3}a) we plot  the monogamy asymmetry of  Gaussian quantum steering in the tripartite system versus the Hawking temperature $T$ for fixed squeezing  $s=1$.  It is shown that  the monogamy asymmetry of steering increases for all the $x$ and $\{yz\}$ bipartite systems with increasing Hawking temperature. This means the thermal bath induced by Hawking radiation destroys the symmetry of steerability monogamy.  We find that  the  steering monogamy is endowed with a maximum asymmetry for some fixed $T$. In addition, the  maximizing  condition for the  monogamy asymmetry is $s=arccosh(\frac{\cosh^2 r}{1-\sinh^2r})$, which is  exactly the transition point of the  $1 \to 1$ steering  asymmetry \cite{wang15} . To explain this coincidence, we calculate the symmetry between the ${1 \to 2}$  steering and   ${2 \to 1}$ steering  and plot them in  Fig. (3b).
In Fig. (3b)   the ${1 \to 2}$  and   ${2 \to 1}$ steerabilities are symmetric for any Hawking temperature $T$ because all the ${1 \to 2}$  steerabilities  equal to their   ${2 \to 1}$ counterparts. Therefore, we arrive at the conclusion that  the asymmetry of  steering monogamy  totally stems from the  $1 \to 1$ steering  asymmetry.

\section{Conclusions}
The effect of the Hawking effect on the  shareability of Gaussian  steering,   monogamy deficit and  monogamy asymmetry of steering in the background of an asymptotically flat black hole are investigated. We defined two types of  monogamy deficits for quantum steering by combining different monogamy inequalities.  For three-mode Gaussian states, the monogamy deficits  acts a quantifier of genuine tripartite quantum  steering.  Then we compared   two types of  monogamy inequalities by defining the monogamy asymmetry.   It is shown that  quantum steerabilities are distributed among different parties due to the influence of Hawking radiation. However, such a distribution is not  free because the monogamy inequalities are still hold in the curved spacetime. In addition, we find that the ``sudden death" point of the $A \to B$ steeing  is in accordance with the ``sudden birth" point of the $\bar B \to B$  steering. This is another powerful evidence for the monogamy of quantum steering  in the curved spacetime.  We demonstrate that the monogamy of quantum steering shows an extreme scenario in the curved spacetime: party $A$ cannot individually steer the parties $B$  and $\bar B$,  but the  collectivity $\{B\bar B\}$  can be steered by the measurements performed on Alice's side. Similar situations exists between the party $\bar B$ and the  collectivity $\{AB\}$. It is worth noting that the  maximizing  condition for the  monogamy asymmetry is  exactly the transition point of the  $1 \to 1$ steering  asymmetry. This  reveals the fact that  the asymmetry of  steering monogamy  totally stems from the  $1 \to 1$ steering  asymmetry because the   ${1 \to 2}$  and   ${2 \to 1}$ steerabilities are  symmetric.

\begin{acknowledgments}
This work is supported by the National Natural Science Foundation
of China under Grant  No. 11675052, No. 11475061, and No. 91536108, the Doctoral Scientific Fund Project of the Ministry of Education of China under Grant No. 20134306120003, and the Postdoctoral Science Foundation of China under Grant No. 2014M560129, No. 2015T80146. 	

\end{acknowledgments}


\end{document}